\documentclass[preprintnumbers,
letter,
superscriptaddress,
prd,nofootinbib,floatfix,12pt,
]{article}

\usepackage{jheppub}

\usepackage{graphicx,amssymb,amsmath,color}
\usepackage{hyperref}
\usepackage{color}
\usepackage{setspace}
\usepackage{enumerate}

\newcommand{\shat}{{\hat s}}
\newcommand{\calA}{{\cal A}}

\def\beq{\begin{equation}}
\def\eeq{\end{equation}}

\def\bea{\begin{eqnarray}}
\def\eea{\end{eqnarray}}
\def\bei{\begin{itemize}}
\def\eei{\end{itemize}}
\def\bmat{\begin{matrix}}
\def\emat{\end{matrix}}
\def\ble{\begin{flushleft}}
\def\ele{\end{flushleft}}
\def\={\,=\,}
\def\+{\,+\,}
\def\-{\,-\,}
\def\GeV{\,{\rm GeV}\,}

\def\fb{\, {\rm fb} \,}

\def\rr{\gamma\gamma}

\def\tb{t_\beta}

\def\br{{\rm Br}}
\def\nn{\nonumber}

\newcommand{\Fig}[1]{Fig.~\ref{#1}}
\newcommand{\Eq}[1]{Eq.~(\ref{#1})}
\newcommand{\Sec}[1]{Sec.~\ref{#1}}
\newcommand{\cb}{c_\beta}
\renewcommand{\sb}{s_\beta}

\title{How Resonance-Continuum Interference Changes 750 GeV Diphoton Excess: \\
Signal Enhancement and Peak Shift}

\author[a]{Sunghoon Jung}
\emailAdd{shjung@slac.stanford.edu}
\affiliation[a]{SLAC National Accelerator Laboratory, Menlo Park, CA 94025, USA}

\author[b]{Jeonghyeon Song}
\emailAdd{jeonghyeon.song@gmail.com}
\affiliation[b]{School of Physics, KonKuk University, Seoul 143-701, Korea}

\author[b,c]{Yeo Woong Yoon}
\emailAdd{ywyoon@kias.ac.kr}
\affiliation[c]{Korea Institute for Advanced Study, Seoul 130-722, Korea}

\abstract{ 
The new scalar resonance contribution to the 750 GeV diphoton excess observed at the LHC 13 TeV necessarily interferes with the continuum background in the $gg\to \gamma \gamma$. The interference has two considerable effects: (1) enhancing or suppressing diphoton signal rate due to the imaginary-part interference and (2) distorting resonance shape due to the real-part interference. From the best-fit study of two benchmark models (two Higgs doublets with $\sim$50 GeV widths and a singlet scalar with 5 GeV width, both extended with vector-like fermions), we find that the resonance contribution to the 750 GeV excess can be enhanced by a factor of 2(1.6) for 3(6) fb signal rate and the 68\%(95\%) CL best-fit mass range can shift by 1--4 (any ${\cal O}(1)$) GeV. If the best-fit excess rate decreases with future data, the interference effects will become more significant. The inevitable interferences can also provide a consistency check of a resonance hypothesis, whether or not future precision shape measurements confirm a Breit-Wigner shape or discover interesting deviations.
}

\preprint{SLAC-PUB-16447}
\begin{document}

\maketitle

%%%%%%%

%%%%%%%%%%%%%%%%%%%
\section{Introduction}

Recently, mild excesses in diphoton invariant mass distribution have been observed in both ATLAS~\cite{atlas:excess} and CMS~\cite{cms:excess} experiments at Large Hadron Collider (LHC) 13 TeV running. The excesses are 3.6$\sigma$ and 2.6$\sigma$ significant from Standard Model (SM) hypothesis, respectively, and are found to prefer a new resonance at around 750 GeV decaying to diphotons~\cite{atlas:excess,cms:excess,Falkowski:2015swt}. The excesses at LHC 13 are currently not completely inconsistent with no significant excesses at LHC 8 TeV data, e.g.~\cite{Falkowski:2015swt}, and more data are needed to confirm or disfavor the resonance interpretation. The  tantalizing hint of a new resonance  triggered various theoretical proposals~\cite{DiChiara:2015vdm,Angelescu:2015uiz,Buttazzo:2015txu,Pilaftsis:2015ycr,Franceschini:2015kwy,Ellis:2015oso,Gupta:2015zzs,Higaki:2015jag, McDermott:2015sck,Low:2015qep,Petersson:2015mkr,Dutta:2015wqh,Cao:2015pto,Kobakhidze:2015ldh,Cox:2015ckc,Martinez:2015kmn,Becirevic:2015fmu,No:2015bsn,Demidov:2015zqn,Chao:2015ttq,Fichet:2015vvy,Curtin:2015jcv,Bian:2015kjt,Chakrabortty:2015hff,Csaki:2015vek,Bai:2015nbs,Benbrik:2015fyz,Kim:2015ron,Gabrielli:2015dhk,Alves:2015jgx,Carpenter:2015ucu,Bernon:2015abk,Chao:2015nsm,Han:2015cty,Dhuria:2015ufo,Han:2015dlp,Luo:2015yio,Chang:2015sdy,Bardhan:2015hcr,Feng:2015wil,Cho:2015nxy,Barducci:2015gtd,Chakraborty:2015jvs,Han:2015qqj,Antipin:2015kgh,Wang:2015kuj,Cao:2015twy,Huang:2015evq,Heckman:2015kqk,Bi:2015uqd,Kim:2015ksf,Cline:2015msi,Bauer:2015boy,Chala:2015cev,Boucenna:2015pav,deBlas:2015hlv,Murphy:2015kag,Hernandez:2015ywg,Dey:2015bur,Huang:2015rkj,Patel:2015ulo,Chakraborty:2015gyj,Altmannshofer:2015xfo,Cvetic:2015vit,Allanach:2015ixl,Cheung:2015cug,Liu:2015yec,Hall:2015xds,Kang:2015roj} allegedly regarded to fit the 750 GeV excess rate $\sim {\cal O}(1)$ fb. Also, both a narrow and a somewhat broad resonance with $\Gamma \sim {\cal O}(10)$ GeV can fit the data similarly well~\cite{atlas:excess,Falkowski:2015swt}.

\begin{figure}[t] \centering
\includegraphics[width=0.6\textwidth]{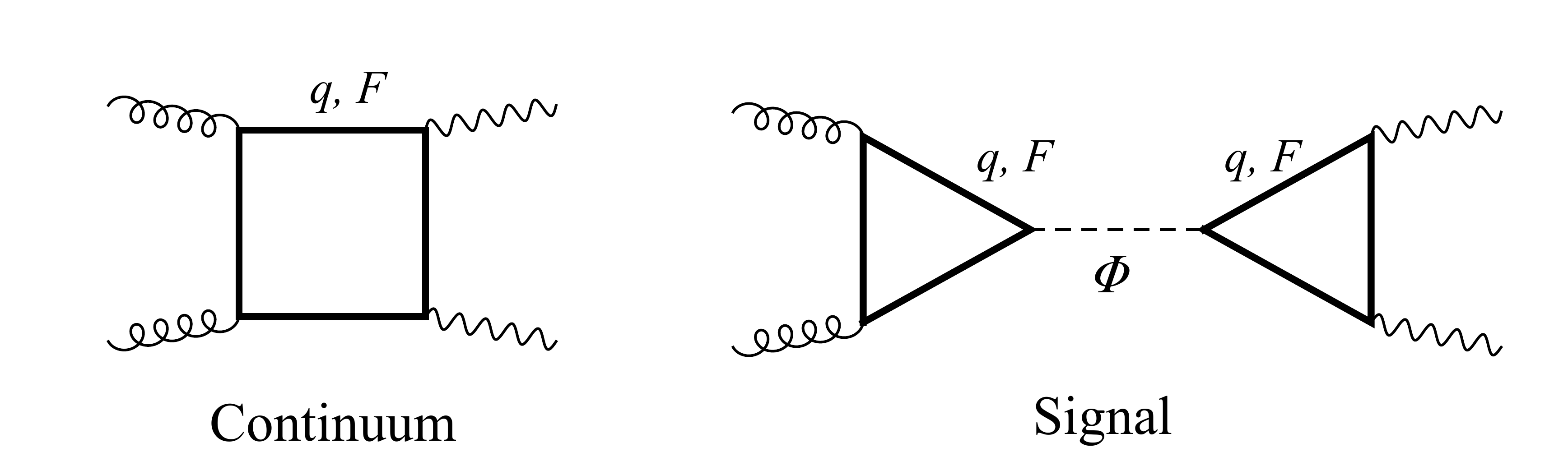}
\caption{
Representative Feynman diagrams of the interfering continuum background (left) and a scalar resonance signal (right) in the $gg \to \gamma \gamma$.
}
\label{fig:feyn:diagram}
\end{figure}

The interference between a resonance and the SM continuum background, however, is inevitable~\cite{Dicus:1987fk,Dixon:2003yb,Martin:2012xc,Dixon:2013haa,Martin:2013ula,Coradeschi:2015tna,Jung:2015gta,Jung:2015sna} but has been ignored so far. The interference can have two considerable effects (see, e.g. Ref.~\cite{Jung:2015gta}): 
\begin{enumerate}
\item Enhancing or suppressing diphoton signal rate,
\item Distorting resonance shape.
\end{enumerate}
The effects can be especially sizable if the resonance width is at least comparable to experimental resolutions or bin sizes, $\Gamma \gtrsim 5$ GeV. For the 125 GeV SM Higgs boson, for example, even though it is narrow,
the resulting peak-shift is $\sim 70$ MeV~\cite{Dixon:2013haa,Martin:2012xc}
and will be comparable to the pole-mass measurement uncertainty soon
(currently $\sim 490$ MeV~\cite{Aad:2015zhl}). 
For a 750 GeV $gg$-fused scalar resonance
with ${\cal O}(1)$ fb diphoton rate, the resonance-continuum interference is generally large: the resonance-squared  $S \sim {\cal O}(1)$ fb and the $gg \to \gamma \gamma$ continuum background $B \sim 0.2$ fb/40 GeV naively generate $2\sqrt{SB}/S \sim  (30-90)$\% relative interference effect. The interference is particularly large in the diphoton channel because the scalar resonance contribution is two-loop suppressed while the interfering continuum background is only one-loop as shown in Fig.~\ref{fig:feyn:diagram}, so that the above naive estimation of the relative interference is generally loop-factor enhanced~\cite{Jung:2015gta,Jung:2015sna}.

The two main interference effects are induced by different relative phases bewteen the resonance and the continuum processes. The real-part interference ($\propto \cos \phi$ as will be defined and discussed) induces either peak-dip or dip-peak pattern added to a resonance peak, hence distorting the resonance shape from a pure resonance peak. On the other hand, the imaginary-part interference ($\propto \sin \phi$) simply rescales the resonance peak, hence enhancing or suppressing the resonance peak. The non-zero phase is generated when some particles running in loops are lighter than 375 GeV.

In this paper, we investigate each interference effect on the current 750 GeV excess data  by considering two benchmark models that exhibit maximally enhanced signals (from the purely imaginary-part interference) or maximally distorted resonance shape (from the purely real-part interference) in the $gg \to \gamma \gamma$ process. We first describe our method of calculating resonance shapes including interferences in \Sec{subsec:formalism} and the diphoton datasets and best-fit analysis method in \Sec{subsec:best-fit}. The two benchmark models are introduced and our main results are discussed in Sec.~\ref{sec:singlet} and \Sec{sec:2hdm}.
Then we conclude and discuss prospects in Sec.~\ref{sec:conclusions}.

%%%%%%%%%%%%%%%%%%%
\section{Formalism and Analysis Method}

%%%%%
\subsection{Diphoton Rate and Resonance Shape}
\label{subsec:formalism}

We consider a scalar resonance in the $gg \to \gamma\gamma$. It interferes with the one-loop continuum backgrounds shown in \Fig{fig:feyn:diagram}.
The total differential cross section including the interference is written as
\bea
\frac{d \sigma}{d m_{\rr}} &=&
 \frac{d \sigma_{\rm bg}}{d m_{\rr}} +
  \frac{d \sigma_{\rm sig}}{d m_{\rr}}  
  \\ \nn
  &=&
 %\frac{d \sigma_{\rm bg}}{d m_{\rr}}
 %+\frac{d \sigma_{\rm sig}}{d m_{\rr}} \=
\frac{2}{m_{\rr}}\, {\cal L}_{gg} \Big(\frac{m_{\rr}^2}{s}\Big)  \Big[ \hat{\sigma}_{\rm bg} (m_{\rr}^2) \,+ \, \hat{\sigma}_{\rm sig} (m_{\rr}^2) \Big],
\label{eq:diffrate}
\eea
where ${\cal L}_{gg}(x)=\int_x^1 dy \, ({x}/{y}) f_{g/p}(y) f_{g/p}(x/y)$ is the $gg$ parton luminosity (we use CT10NNLO PDF set~\cite{Gao:2013xoa}) and $\hat{\sigma}_{\rm bg, sig}$ are 
the parton-level cross sections.
The signal cross-section $\hat{\sigma}_{\rm sig}$, the deviation from the SM background,
consists of the resonance-squared and the resonance-continuum interference~\cite{Jung:2015gta},
\bea
\hat{\sigma}_{\rm sig}(\hat{s}) &=& \frac{M^4}{(\hat{s}-M^2)^2 + M^2 \Gamma^2 } \, \left[ \left( 1+\frac{2\Gamma}{RM} s_\phi \right) \hat{\sigma}_{\rm res}
+ \frac{2(\hat{s} -M^2)c_\phi }{M^2} \hat{\sigma}_{\rm int} \right],
\label{eq:shape}
\eea
where $s_\phi=\sin \phi$ and $c_\phi=\cos \phi$,
and we factor out Breit-Wigner (BW) parts.
We define $\hat{\sigma}_{\rm res, int}$ and the relative phase $\phi$ in terms of phase-space integrated squared amplitudes (${\cal M}_i={\cal A}_i e^{i\phi_i}$)
\bea
\label{eq:phi:definition}
{\hat \sigma}_{\rm int} e^{i\phi} &\equiv&
\frac{1}{32\pi\shat} \int d\cos\theta^*
\sum \calA_{\rm bg}\calA_{\rm res} e^{i(\phi_{\rm res}
- \phi_{\rm bg})},\eea
and similarly for $\hat{\sigma}_{\rm res}$. The summation is over helicity and color indices, and $\theta^*$ is the scattering angle in the c.m.~frame.
We introduce a key parameter $R$, defined by
\bea
\label{eq:candr}
R &\equiv& \frac{{\hat \sigma}_{\rm res}}{{\hat \sigma}_{\rm int}}\approx
\frac{{\cal A}_{\rm res}}{ {\cal A}_{\rm bg} },
\eea
which measures the relative size of interference.

For a narrow resonance (whose width is not much larger than experimental resolutions or bin sizes), the real part interference,
the term proportional to $c_\phi$ in Eq.~(\ref{eq:shape}), is washed out after the integration over $m_{\rr}$. Since the invariant mass distribution is highly accumulated near the resonance peak, we can consider parameters $R$, $\phi$ and parton luminosity as constant values.
Then the total signal rate with the interference effect, defined as $\sigma_{\rm mNWA}$, is obtained as~\cite{Jung:2015gta}
\begin{eqnarray}
\sigma_{\rm mNWA} &=& \sigma_{\rm NWA} \cdot C =
\bigg[\frac{M\pi}{\Gamma} {\cal L}_{gg}\big(M^2/s\big)
\hat{\sigma}_{\rm res} (M^2) \bigg]
 \cdot C ,
\, \, \, 
 \hbox{for a narrow resonance,} \quad
\label{eq:mnwa}
\end{eqnarray}
where $C = (1+ \tfrac{2\Gamma}{RM} s_\phi )$ quantifies the strength of the imaginary-part interference.
Note that the terms inside the square bracket
corresponds to the usual total rate in the narrow width approximation (NWA),
production cross section times branching ratio. The subscript mNWA represents {\it modified} NWA. 
It is useful to express $\hat{\sigma}_{\rm sig}(\hat{s})$ in terms of  $\sigma_{\rm mNWA}$ which is measured in experiments:
\bea
\hat{\sigma}_{\rm sig}(\hat{s})
&= &  \frac{\Gamma M^3/ (\pi {\cal L}_{gg}(M^2/s))}{(\hat{s}-M^2)^2 + M^2 \Gamma^2 } \, \left[ \frac{2(\hat{s} -M^2)}{M^2} \frac{c_\phi}{RC} + 1\right]
\cdot \sigma_{\rm mNWA}\,.
\label{eq:narrowshape}
\eea
This is our resonance shape function for a narrow resonance.

 For a broad resonance, with $\Gamma \gtrsim 50$ GeV, we now need to take into account the  $m_{\rr}$ dependence of $R$, $\phi$ and parton luminosity; they are not constant in $m_{\rr}$ anymore in broad resonance region.
 We redefine the total rate $\sigma_{\rm mNWA}$ for a broad resonance by integrated differential rate, Eq.~(\ref{eq:diffrate}), around the resonance mass $M$:
 \begin{equation}
 \sigma_{\rm mNWA} = \int_{M-\Delta}^{M+\Delta}dm_{\rr}
 \bigg[ \frac{d\sigma_{\rm sig}}{dm_{\rr}}  \bigg]_{\rm peak},
 \quad \hbox{for a broad resonance}.
 \end{equation}
 We set $\Delta = 100\GeV$ for our broad resonance example. We also use the following ratio
 \begin{equation}
 \label{eq:Kintf}
 K_{\rm intf} = \frac{\sigma_{\rm mNWA}}{ \sigma_{\rm prod}\cdot \br_{\rr}}
 \end{equation}
 to quantify the strength of the imaginary-part interference for a broad resonance.
This $K_{\rm intf}$ factor is approximately equal to the $C$ factor for a narrow resonance in \Eq{eq:mnwa}.

\medskip
The resonance shape function
 is parameterized not  only by usual mass $M$, width $\Gamma$ and the total rate $\sigma_{\rm mNWA}$ but also by the relative interference phase $\phi$.
  $R$ is not a completely independent parameter as shall be discussed.
The purely real-part (imaginary-part) interference corresponds to 
$\phi=0,180^\circ $ ($\phi=\pm 90^\circ$). The real-part interference induces peak-dip or dip-peak structure in addition to a BW peak 
while the imaginary-part interference either enhances or reduces the BW peak or convert the peak to a BW dip (without associated peak)~\cite{Jung:2015gta}. Thus, the purely real-part interference can most significantly change the resonance shape from a BW peak while the purely imaginary-part interference can most significantly enhance the signal rate (or peak height). These two effects are our main topics; we will study two benchmark models for each of them.

It is hard to carry out a model-independent best-fit analysis including interference effects based on \Eq{eq:shape} and \Eq{eq:narrowshape}. The interference depends not only on $M$, $\Gamma$, $\sigma_{\rm mNWA}$,
which  are usually chosen in model-independent analysis without interference effects, but also on $\phi$ and $R$. In particular, $R$ is correlated with $\sigma_{\rm mNWA}$,
which is hard to obtain the analytic relation. In this regard
 we use two benchmark models to numerically discuss the interference effects.
For the (purely) real-part interference, we consider a singlet model
which introduces a CP-odd SM singlet scalar with a minimal set of vector-like quarks and vector-like leptons: see \Sec{sec:singlet}.
For the (purely) imaginary-part interference, Type II 2HDM
with vector-like leptons is to be studied: \Sec{sec:2hdm}.

There is an important assumption in our implementation of higher-order corrections. We first normalize the total rate without interferences (equivalent to multiplying the correction factor to $\hat{\sigma}_{\rm res}$) to the result obtained by \texttt{HIGLU} fortran package~\cite{Spira:1995mt} which includes next-to-next-to-leading-order QCD and next-to-leading-order EW contributions.  Then we multiply the same correction factor to the interference term $\hat{\sigma}_{\rm int}$, as no results are available.
Although this assumption approximately accounts for higher-order corrections to the total rate, it implies that $R=\hat{\sigma}_{\rm res}/\hat{\sigma}_{\rm int}$ does not receive appreciable higher-order corrections.
This may not be an unreasonable assumption since higher-order corrections to the resonance-squared and the resonance-continuum interference can be similar, hence cancelling out in their ratio $R$. In any case, both the purely real-part and the purely imaginary-part interferences approximately grow with $1/R$. Thus, any corrections to $R$ would directly affect what we discuss in this paper.

%%%%%
\subsection{Dataset and Method} 
\label{subsec:best-fit}
In order to quantitatively study interference effects on the 750 GeV diphoton excess data, we perform a Poissonian likelihood analysis to find the best fit.
The dataset is from the latest LHC 8 and 13 TeV diphoton resonance search data at around $m_{\gamma \gamma} = 750$ GeV from both ATLAS and CMS experiments. We read in the predicted backgrounds and observed data from the reported plots in Refs.~\cite{atlas:excess,cms:excess,Aad:2015mna,Khachatryan:2015qba}. The total uncertainty in each bin is assumed to be $2\,(1.5) \, \times$ statistical uncertainty for LHC 13\,(8) TeV data. 

The fit ranges considered in this paper are 
\bea
m_{\gamma \gamma} &=& \{ 630, \, 830 \} \, {\rm GeV} \quad \textrm{for ATLAS 13 (3.2/fb), CMS 13 (2.6/fb), CMS 8 (19.7/fb)}, \nonumber\\
m_{\gamma \gamma} &=& \{ 642, \, 835 \} \, {\rm GeV} \quad \textrm{for ATLAS 8 (20.3/fb)}.
\eea
We choose ATLAS 8 data bins closest to 630 and 830 GeV. The range is somewhat broad so that we can consider a broad resonance as well. CMS 13 dataset is divided into CMS EBEB 13 and CMS EBEE 13 categories depending on which parts of detectors identify photons. We consider them as independent datasets. Fiducial signal efficiencies are taken from the experimental references and Ref.~\cite{Falkowski:2015swt}.

We carry out a $\chi^2$-fit to all the data bins within the range, and take the total change of $\chi^2$ compared to the SM-fit (background-only), $\Delta \chi^2 = \chi^2 - \chi^2_{\rm SM}$, as a measure of how well the model fits the data. 
Our SM-fit (background-only) results are: 
\beq
\chi^2_{\rm SM} = 7.02, \, 4.93, \, 17.77, \,  1.52, \, 16.65,
\eeq
for ATLAS 13, CMS EBEB 13, CMS EBEE 13, ATLAS 8, and CMS 8, respectively. The results are, of course, sensitive to the assumption of total uncertainties. As will be discussed, although CMS EBEE 13 and CMS 8 show worst fits, these data do not strongly support a 750 GeV resonance -- various excesses and deficits around 750 GeV are not significantly fitted better with new resonance contributions. However, ATLAS 13 and CMS EBEB 13 data are fitted better with a new resonance at around 750 GeV.
Our read-in data and model-independent fit results without interferences approximately agree with those in Ref.~\cite{Falkowski:2015swt}; assuming a BW peak with both fixed $\Gamma=5,\,40$ GeV and with varying $\Gamma$.

%%%%%%%%%%%%%%%%
\section{Singlet Model: Real-Part Interference} \label{sec:singlet}

%%%%%
\subsection{Singlet Model}
Consider a CP-odd SM-singlet scalar $\Phi=A$, coupling to vector-like quarks $Q \equiv Q^{7/6}=({\bf 3},{\bf 2},7/6)$ and vector-like leptons $L \equiv L^{3/2}=({\bf 1},{\bf 2},3/2)$
\beq
{\cal L} \, \ni \, \frac{1}{2}M_\Phi^2 \Phi^2 \, + \, \sum_Q (s_Q \Phi + M_Q) \overline{Q} \gamma_5 Q \, + \, \sum_L ( s_L \Phi + M_L) \overline{L} \gamma_5 L,
\eeq
%and similarly for a CP-even SM-singlet scalar $\Phi=H$,
%\beq
%{\cal L} \, \ni \, \frac{1}{2}M_\Phi^2 \Phi^2 \, + \, \sum_Q (s_Q \Phi + M_Q) \overline{Q} Q \, + \, \sum_L ( s_L \Phi + M_L) \overline{L} L,
%\eeq
where $s_{Q,L}$ are real Yukawa couplings, $M_{\Phi, Q, L}$ mass eigenvalues, $N_{Q,L}$ number of fermions, and $q_{Q,L}$ electric charges. 
We choose $Q^{7/6}$ and $L^{3/2}$ from the minimal matter list~\cite{DelNobile:2009st} -- the list of new particles that can eventually decay to SM particles -- since they have the largest electric charges. We consider $A$, but $H$ shall also exhibit similar effects.

In the quark sector, we introduce a single vector-like $Q$ with fixed parameters
\beq
M_Q=1\, {\rm TeV}, \, N_Q = 2, \, s_Q=0.2.
\eeq 
We still have enough lepton sector free parameters that we can use to fit the data and to illustrate interference effects.

In the lepton sector, we consider
\beq
M_L = 400 \, {\rm GeV}, \, N_L = 6, \, s_L \textrm{ is varied}.
\eeq
The sign of the Yukawa $s_L$ determines the sign of the relative phase: $s_L \to -s_L$ approximately changes the relative phase $\phi \to \pi + \phi$. It is an approximate relation because $Q$ also contributes to the $\Phi \to \gamma \gamma$ part although it is subdominant to the $L$ contribution. We will compare the results with positive and negative $s_L$ (as well as with the results without any interference accounted for) to see how the best-fit changes with interference effects.

Another important parameter is the width. In the above model, the width is typically too small ($\lesssim 1$ GeV) to make interference effects apparent in current experiments; $\Phi$ mainly decays to loop-induced $gg$ and $\gamma \gamma$ 
\bea
\Gamma(\Phi \to gg ) &=& \frac{\alpha_S^2 }{128 \pi^3} \frac{M_\Phi^3}{M_Q^2} \left| \sum_Q s_Q A_{1/2}^\Phi \left( \frac{M_\Phi^2}{4 M_Q^2} \right) \right|^2, \\
\Gamma(\Phi \to \gamma \gamma ) &=& \frac{\alpha^2 }{256 \pi^3} M_\Phi^3 \left| \sum_{f=Q,L} N_C \, q_f^2 \frac{s_f}{M_f} A_{1/2}^\Phi \left( \frac{M_\Phi^2}{4 M_f^2} \right) \right|^2, 
\eea
where loop functions $A_{1/2}^{\Phi=H,A}$ are defined as in Ref.~\cite{Djouadi:2005gi}, and other signals such as $Z\gamma,\, ZZ,\, WW$ are currently well below their LHC 8 sensitivities. 
If such a narrow resonance falls within a single experimental bin, the real-part interference (although itself is independent on the width) is cancelled out; in addition, the imaginary-part interference is small since it is directly proportional to the width as $C-1 \propto \Gamma$. Thus, to illustrate possible impacts of interference effects, we assume a bigger constant width
\beq
\Gamma_\Phi = 5 \, {\rm GeV}.
\eeq
It is easy to add extra hidden decay modes of $\Phi$, not constrained at all, to make so. If the assumed width were much bigger than the true width, diphoton signal will be suppressed; but if the true width is bigger, the interference will become more relevant. Also, if the $N_L$ were smaller, although one can still have almost 100\% BR($\gamma \gamma$), the total width decreases and the interference effects will be less significant. Meanwhile, for $M_L \leq M_\Phi/2$, the decays into vector-like leptons dominate and the diphoton signal becomes too suppressed. Although such light leptons can change the phase $\phi$ and introduce different interference effects, we cannot fit the diphoton excess data well and do not discuss this possibility further. 

An important feature of the singlet scalar model is that the relative phase 
is small:
\beq
\phi \simeq 
\left\{
\begin{array}{ll}
8.3^\circ & \hbox{ for } s_L>0; \\
188.3^\circ & \hbox{ for } s_L<0, 
\end{array}
\label{eq:phi}
\right.
\eeq
which induces almost purely real-part interference. This is the case in which resonance shape is maximally distorted from pure BW shape (and the peak location is maximally shifted), for the given total rate. The small but non-zero phase is generated from the SM quark loops in $gg \to \gamma \gamma$ background box diagrams.

%%%
\subsection{Results -- Singlet Model} \label{sec:result-singlet}

\begin{figure}[t] \centering 
\includegraphics[width=0.6\textwidth]{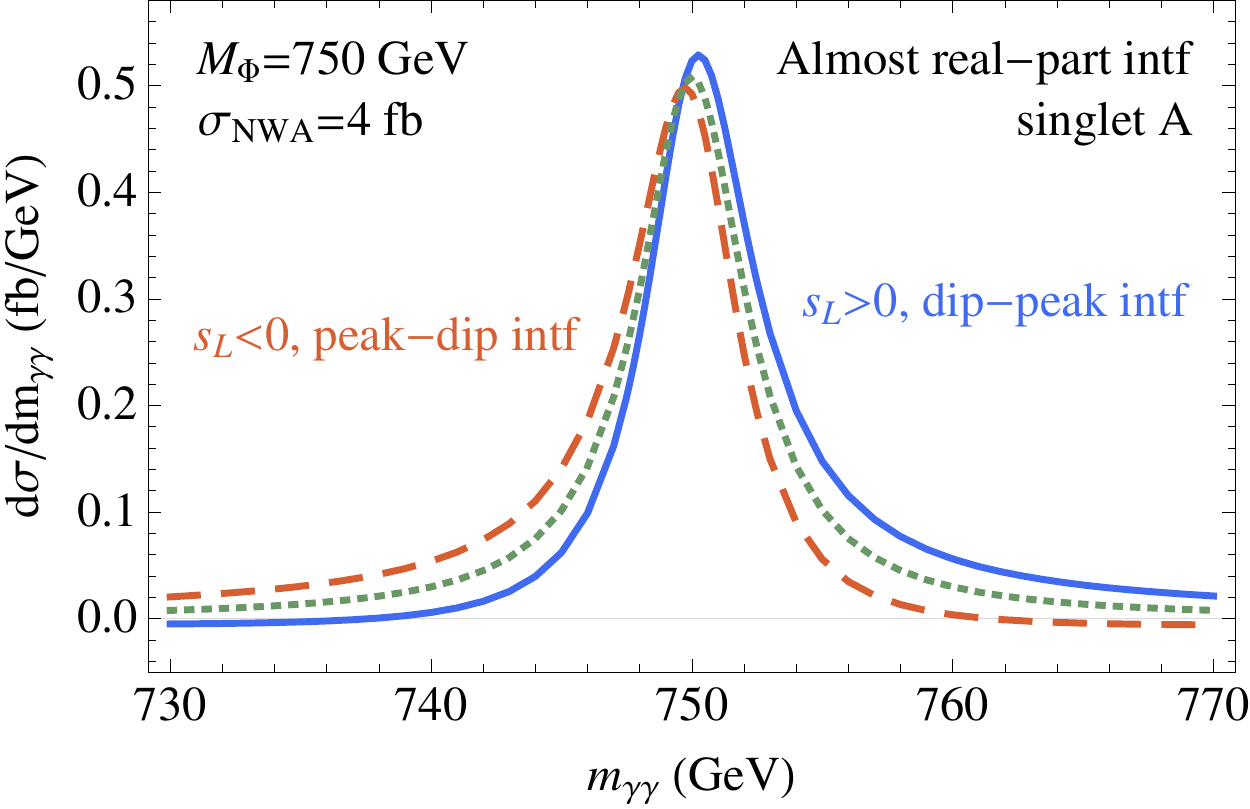}
\caption{
Example diphoton resonance shapes with $s_L>0$ (dip-peak interference, {\bf blue-solid}),  $s_L<0$ (peak-dip interference, {\bf red-dashed}), and no interference ({\bf green-dotted}) for the same mass $M_\Phi=750$ GeV and the NWA rate $\sigma_{\rm NWA} \simeq 4$ fb. The relative phase $\phi \simeq 8.3^\circ (+ \pi)$ for $s_L>0 (<0)$ induces almost purely real-part interference, and the resulting peak shifts and long tails affect best-fit analysis. The small imaginary-part interference also makes true observable mNWA rates $\sigma_{\rm mNWA}$ and peak heights slightly different. 
We set $|s_L| \simeq 1.5$ and $\Gamma_\Phi =$5 GeV.
}
\label{fig:shape}
\end{figure}

In \Fig{fig:shape} we show an example of the SM-singlet scalar resonance shapes
for $s_L>0$ (blue-solid) and $s_L<0$ (red-dashed) with full interference effects.
For comparison, we also show the resonance shape without any interferences taken into account (green-dotted).  All three cases have the same NWA rates and the width $\Gamma_\Phi$. But $s_L>0 \,(<0)$ induces a small dip-peak (peak-dip) interference pattern added to the BW peak, so that a long tail toward a high (low) invariant mass region appears and the peak shifts toward the same direction. As a result, the best-fit results change, even though the NWA rates, masses and widths are all the same. We quantify such interference effects in this subsection. The small but non-zero imaginary-part interference, \Eq{eq:phi}, actually makes $\sigma_{\rm mNWA}$ (true observable rate slightly different from the NWA rate; see \Eq{eq:mnwa}) and the peak heights slightly different among the three shapes.

\begin{figure}[t] \centering 
\includegraphics[width=0.49\textwidth]{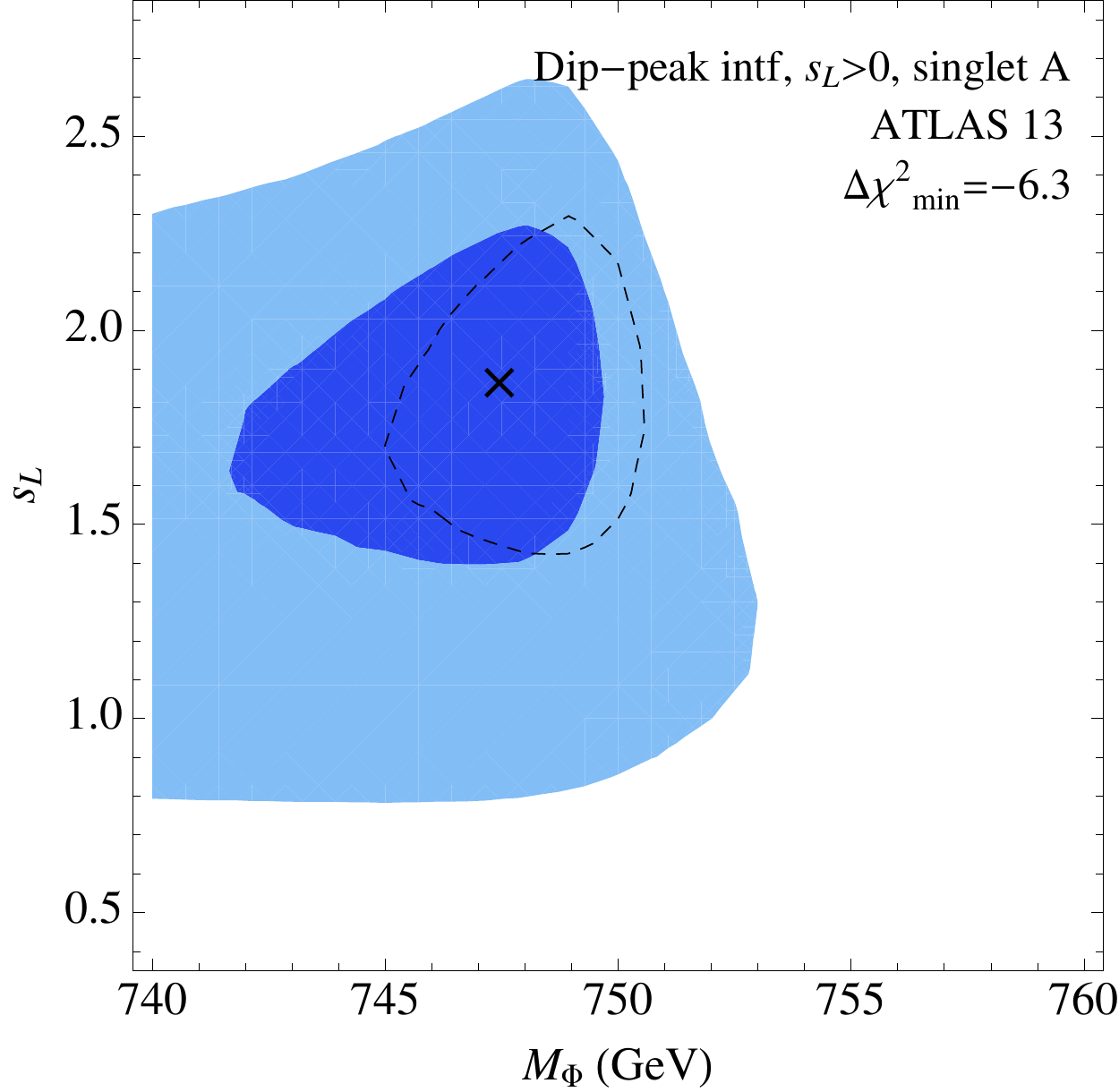}
\includegraphics[width=0.49\textwidth]{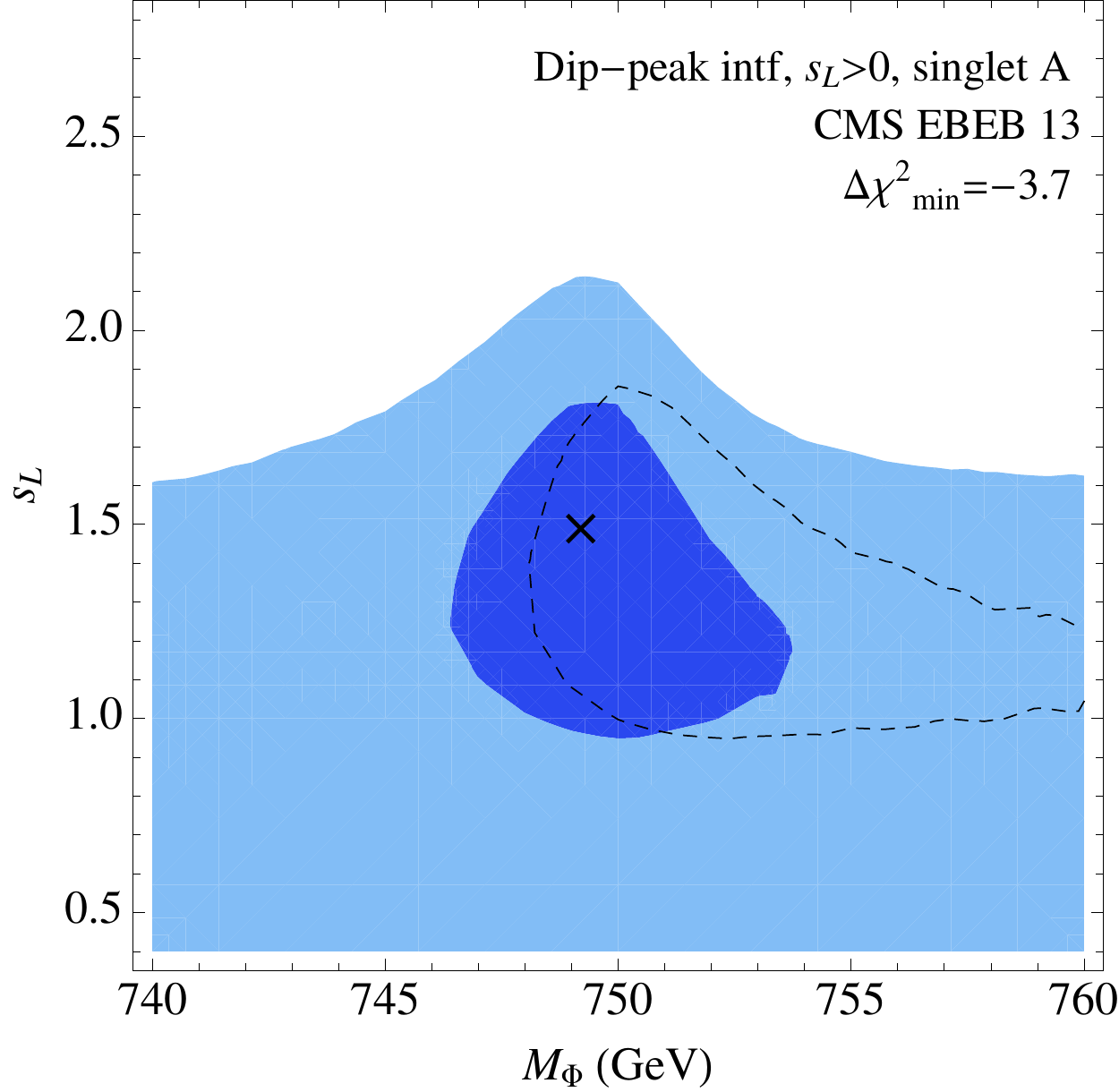}\\
\includegraphics[width=0.49\textwidth]{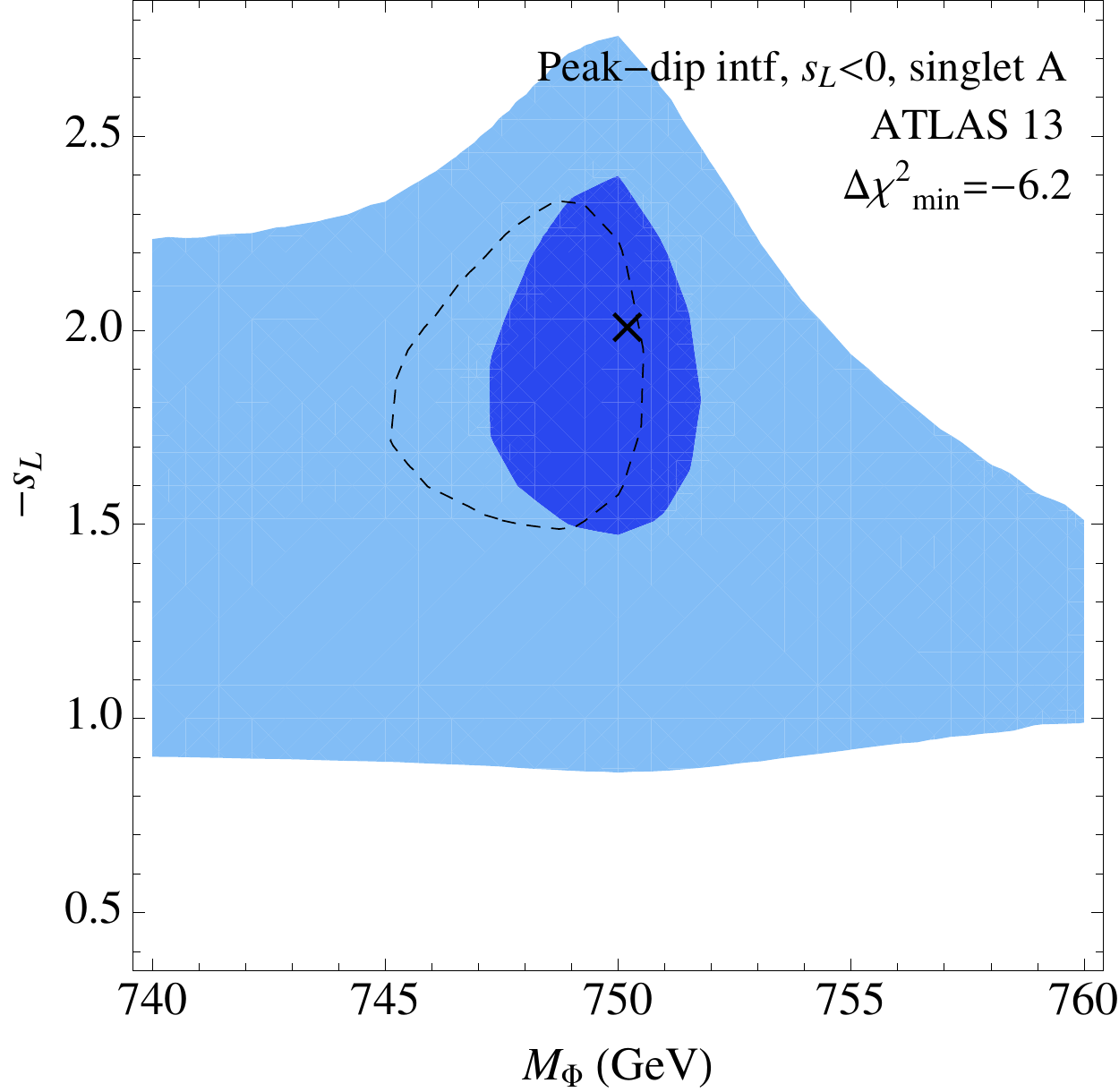}
\includegraphics[width=0.49\textwidth]{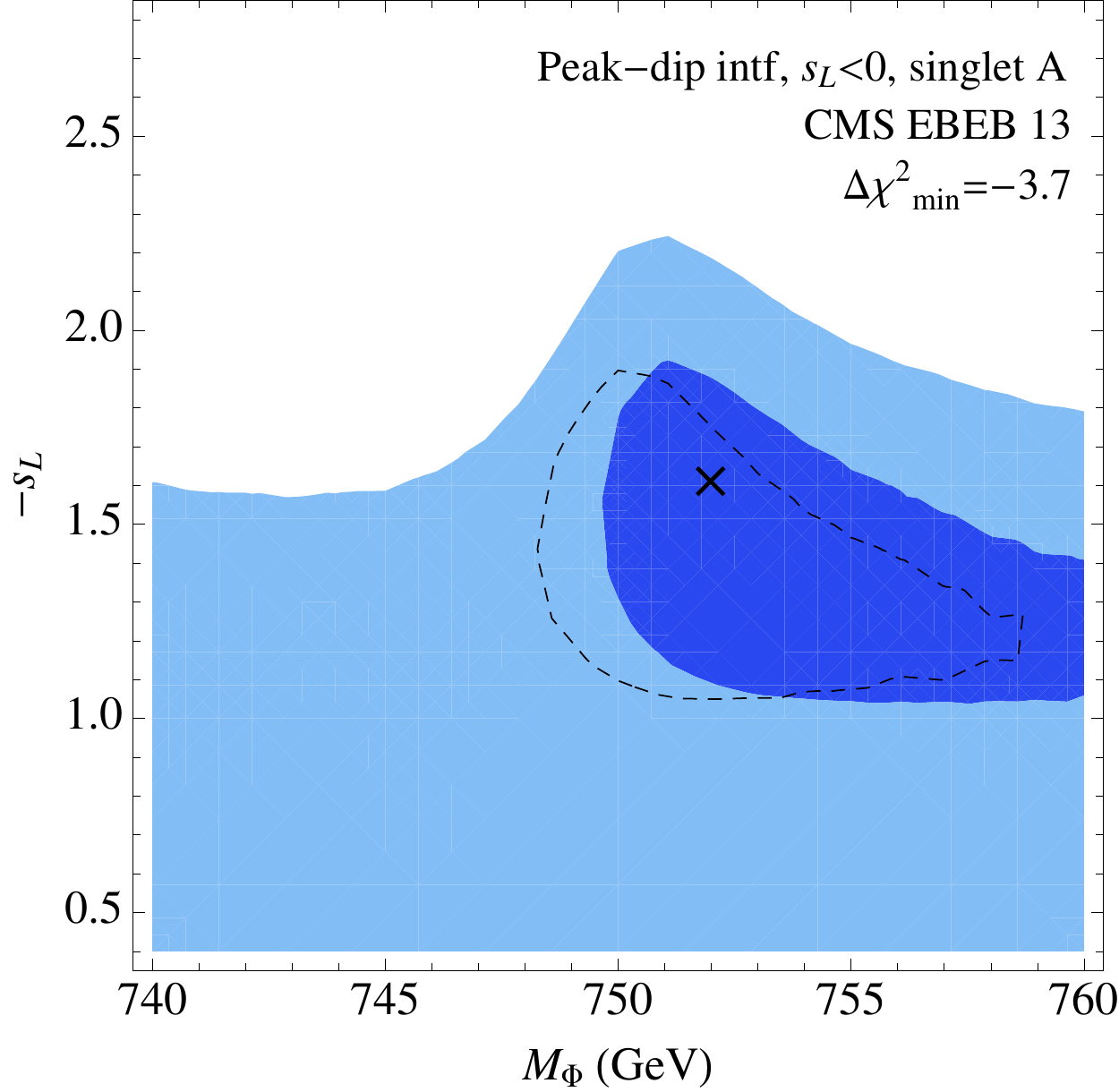}
\caption{
The 68\% CL(darker blue) and 95\% CL(lighter blue) preferred regions for CP-odd singlet $A$. $s_L >0$ ({\bf upper}) and $s_L<0$ ({\bf lower}) can be compared with each other (and with dashed lines for the 68\% CL results without interferences accounted for) to see interference strength. $\Gamma_\Phi$=5 GeV. Fit is performed for $m_{\gamma \gamma} = 630-830$ GeV from ATLAS 13 ({\bf left}, $\chi^2_0=7.02$) and CMS EBEB 13 ({\bf right}, $\chi^2_0=4.93$) datasets.  The best-fit mass ranges with positive and negative $s_L$ differ by ${\cal O}(1)$ GeV and the difference is bigger with weaker $s_L$. The best-fit $\Delta \chi^2_{\rm min}$ compared to the SM fit $\chi^2_0$ is also shown. 
\bigskip \bigskip \bigskip \bigskip
}
\label{fig:resonance13}
\end{figure}
\begin{figure}[t] \centering 
\includegraphics[width=0.49\textwidth]{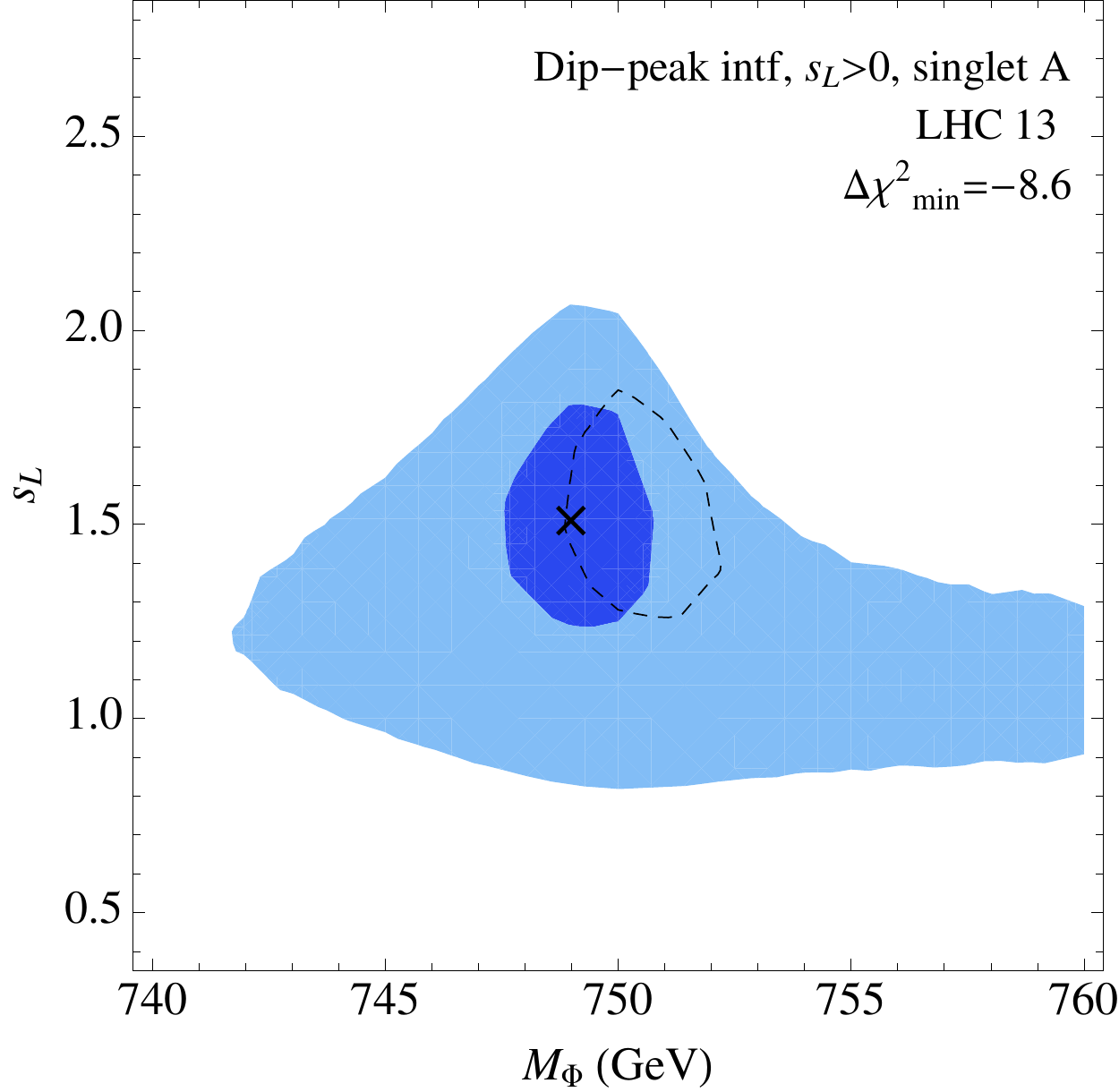}
\includegraphics[width=0.49\textwidth]{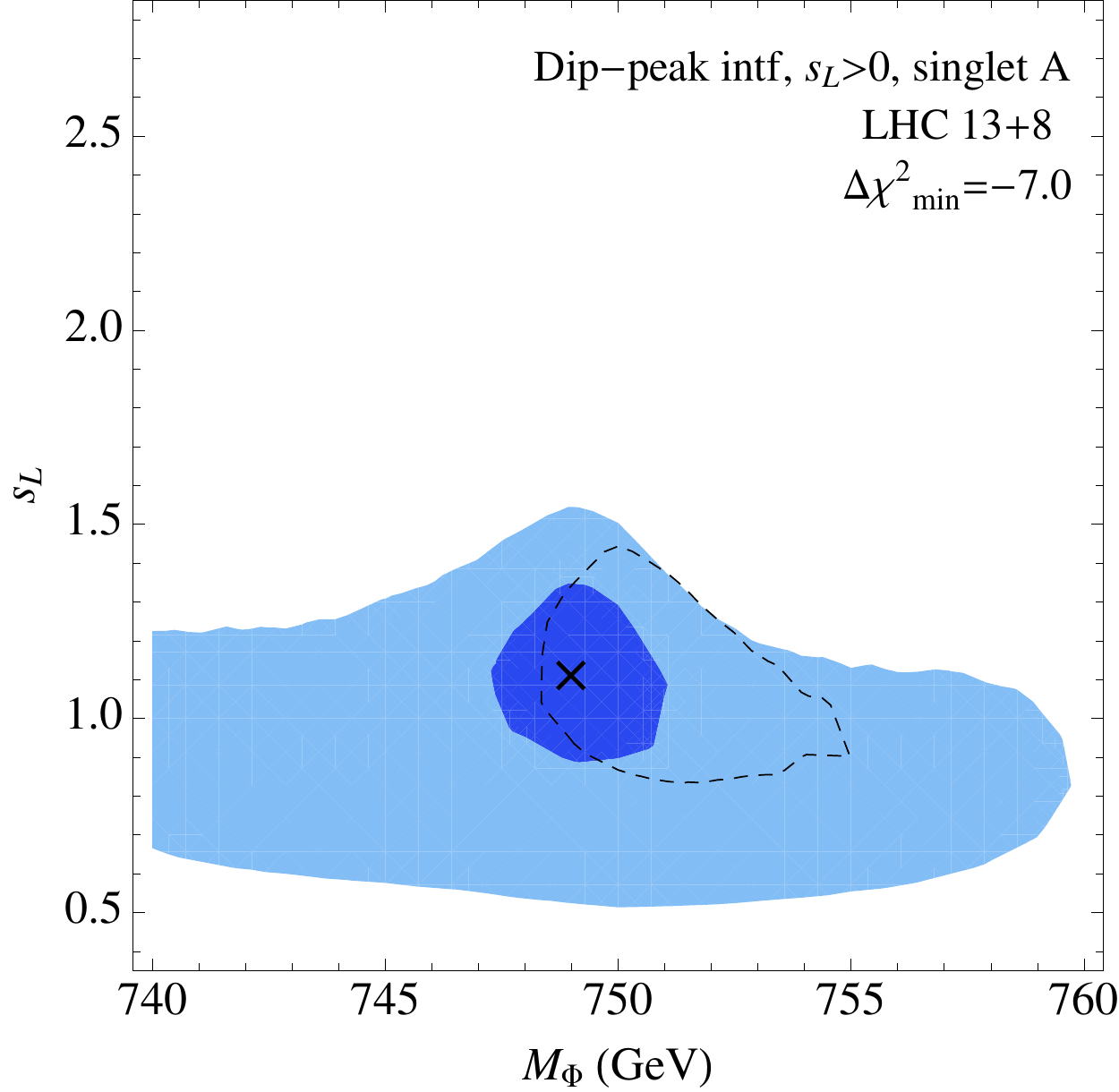}\\
\includegraphics[width=0.49\textwidth]{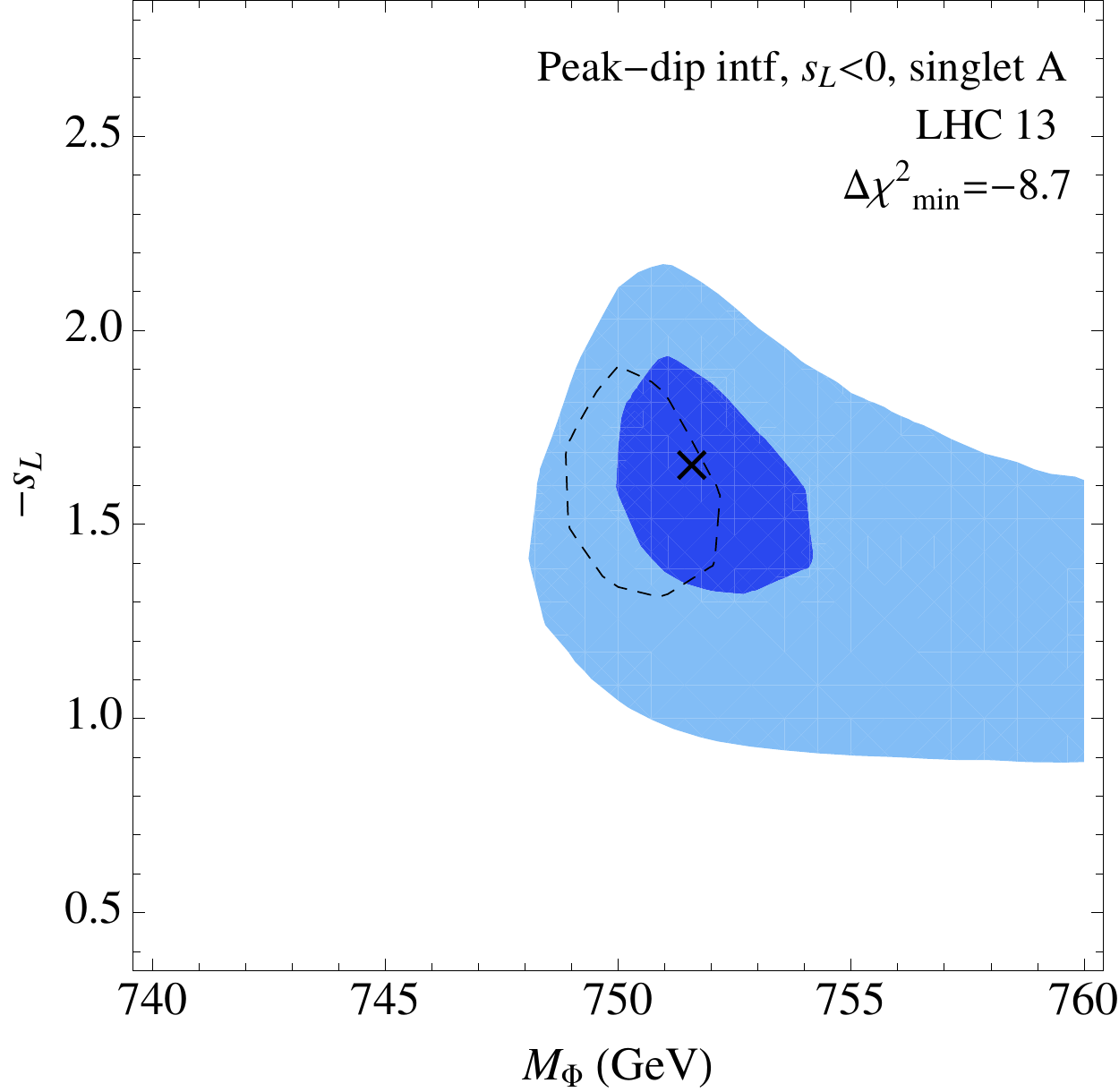}
\includegraphics[width=0.49\textwidth]{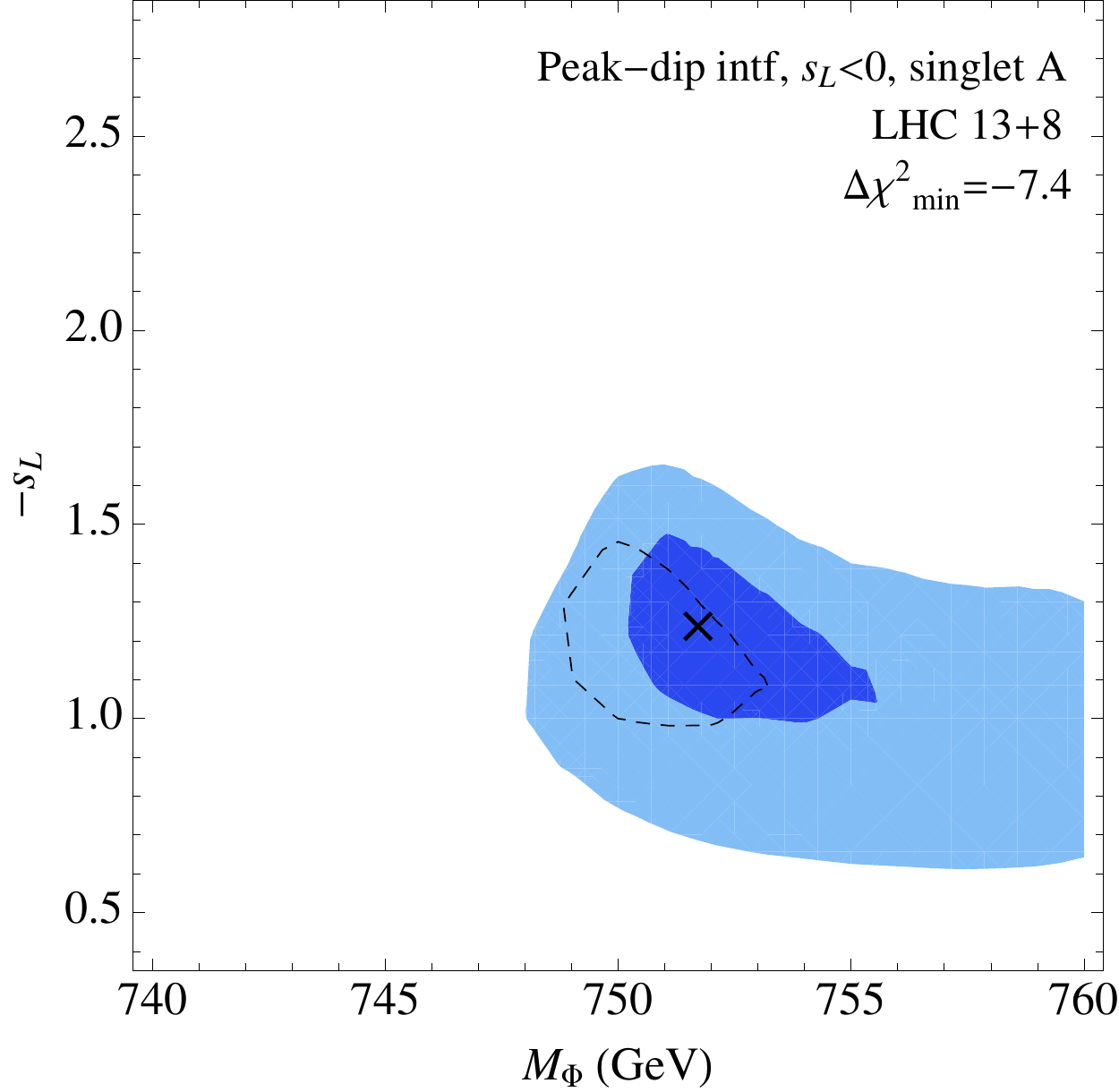}
\caption{The 68\% CL(darker blue) and 95\% CL(lighter blue) preferred regions for CP-odd singlet $A$. $s_L >0$ ({\bf upper}) and $s_L<0$ ({\bf lower}) can be compared with each other (and with dashed lines for the 68\% CL results without interferences accounted for) to see interference strength. $\Gamma_\Phi$=5 GeV. Fit is performed for $m_{\gamma \gamma} = 630-830$ GeV from LHC 13 ({\bf left}, $\chi^2_0=29.7$) and LHC 13+8 ({\bf right}, $\chi^2_0=47.9$)  datasets.  The best-fit mass ranges with positive and negative $s_L$ differ by ${\cal O}(1)$ GeV and the difference is bigger with weaker $s_L$. The best-fit $\Delta \chi^2_{\rm min}$ compared to the SM fit $\chi^2_0$ is also shown. 
\bigskip \bigskip \bigskip \bigskip
}
\label{fig:all13+8}
\end{figure}

\Fig{fig:resonance13} shows the best-fit results to ATLAS 13 (left) and CMS EBEB 13 (right) datasets individually, for a singlet scalar $\Phi=A$ model with $s_L>0$ (upper)
and $s_L<0$ (lower). For comparison, we also show the results without any interferences accounted for (dashed). These datasets are the ones that most strongly prefer the existence of a 750 GeV resonance, and the interference effect does not change the preference of the resonance existence; the data fit much better with a new resonance around 750 GeV even with interferences. Comparing the upper panels for $s_L >0$ with the lower panels for $s_L <0$, we find that the 68\% CL best-fit mass parameter is shifted by about 1--4 GeV while a much bigger shift ${\cal O}(1)$ GeV is expected for the 95\% CL region or for weaker couplings $s_L$. Meanwhile, similar magnitudes of couplings are preferred regardless of interference effects. For $s_L>0$ with dip-peak interference, the peak shifts toward high-mass region and the high-mass region is more accumulated (see \Fig{fig:shape}); consequently, somewhat smaller masses are preferred compared to the $s_L<0$ case (and to the case without interferences).

The interference effects are still apparent, even after including all other LHC 13 and LHC 8 datasets that do not strongly prefer the existence of an additional resonance. This is shown in \Fig{fig:all13+8}; the 68\% CL best-fit regions again shift by about 1--4 GeV and a bigger shift is expected for the 95\% CL region or for weaker couplings $s_L$. The preference of an additional resonance also still exists with interference effects. 

There is a noticeable tendency that interference effects become stronger with a weaker $s_L$, as can be deduced from a wider best-fit mass shift with a weaker $s_L$ in \Fig{fig:resonance13} and \Fig{fig:all13+8}. This is a general result of interference; the real-part interference approximately grows with $1/R \sim {\cal A}_{\rm bg} / {\cal A}_{\rm res}$ amplitude ratio, which measures the background-resonance interference contribution compared to the resonance-squared contribution. If future data prefer to a weaker signal, the interference effects will be larger and more important.

Finally, we briefly compare best-fit results to various datasets. that compared to the ATLAS 13 result in \Fig{fig:resonance13}, the CMS EBEB 13 prefers to a resonance with a slightly higher mass and weaker coupling. But the preferences of a new resonance around 750 GeV from both data are consistent with each other. Including LHC 8 datasets in \Fig{fig:all13+8} significantly prefers to a weaker coupling and actually worsens the best-fit (total $|\Delta \chi^2_{\rm min}|$ in the right panel decreased from the left panel). This may imply that the LHC 8 datasets do not strongly favor the resonance contribution. Future data can only clarify the origin of the excess.

%%%%%%%

\section{VLL-2HDM: Imaginary-Part Interference} \label{sec:2hdm}
\subsection{VLL-2HDM Model}
We consider the Type II two-Higgs-Double-Model (2HDM) in the alignment limit extended with extra vector-like leptons (VLL). We first summarize the (heavy) Higgs sector and then introduce the VLL sector. 

The Higgs sector consists of three neutral Higgs bosons, $h$, $H$ (scalar), $A$ (pseudo-scalar) and two charged Higgs bosons $H^\pm$. In the alignment limit, the $h$ is the 125 GeV SM Higgs boson, and the heavier Higgs bosons, $H$ and $A$, are our focus in this paper. To be consistent with electroweak precision data and to explain the diphoton excess, we consider a degenerate heavy Higgs bosons
\bea
M_\phi = M_H=M_A =750\GeV.
\eea
In the alignment limit with small $t_\beta \sim 1$ as we will focus in this paper\footnote{The parameter space of $\tb \gtrsim 25 $ at $M_{\phi}=750\GeV$ is excluded by the 8 TeV LHC data~\cite{large:tb:exclusion:LHC}.}, the 2HDM alone cannot explain the diphoton excess. It is mainly because the heavy Higgs bosons dominantly decay to the top pair (the decays to $ZZ$ and $WW$ are forbidden in the alignment limit), and the relevant the diphoton branching ratio is only $\br(\phi \to \rr)=7.8\,(8.7) \times 10^{-6}$ for $\phi=H(A)$, leading to too small signal rates
$\sigma(pp\to H,A \to \rr)=0.012\fb$. Thus, we extend the model by extra VLL to achieve the needed $\sim$400 enhancement of diphoton signal.

\begin{table}[t]
\begin{center}
\begin{tabular}{|c|c|c|}
\hline
\hline
 & &~~ ${\rm SU(3)}\times{\rm SU(2)}\times{\rm U(1)}_Y$  \\
\hline
~~$L_L = \left( \begin{array}{c} E_L \\ D_L \end{array} \right)$~~
& ~~$L_R = \left( \begin{array}{c} E_R^{\prime} \\ D_R^{\prime} \end{array} \right)$~~
%& (\textbf{3}, \textbf{2}, $\tfrac{1}{6}$)
& (\textbf{1}, \textbf{2}, $-\tfrac{3}{2}$)\\
$E_R$ & $E_L^{\prime}$
%& (\textbf{3}, \textbf{1}, $\tfrac{2}{3}$)
& (\textbf{1}, \textbf{1}, $-1$)\\
$D_R$ & $D_L^{\prime}$
%& ~(\textbf{3}, \textbf{1}, $-\tfrac{1}{3}$)~
& ~(\textbf{1}, \textbf{1}, $-2$)~\\
\hline
\hline
\end{tabular}
\caption{
 \baselineskip 3.5ex \label{tab:Q.no}
 The contents and quantum numbers of vector-like leptons in the VLL-2HDM model. The electric charges of the doublet components are $(-1,-2)$. }
 \end{center}
\end{table}

We now introduce VLLs,
$L_L$, $E_R$, $D_R$, $E_L^{\prime }$, $D_L^{\prime }$, 
of which the quantum numbers are summarized in Table \ref{tab:Q.no}.
Note that the electric charges of $E^{(\prime)}$
and $ D^{(\prime)}$ are 
$-1$ and $-2$, respectively. 
All of the VLLs in Table \ref{tab:Q.no} are imbedded in one family. 
The Lagrangian of the VLLs in Type II 2HDM is
\begin{eqnarray}
-{\cal L} &=&
Y_D \overline{L}_L H_1 D_R
+Y_D^\prime \overline{L}_R H_1 D_L^{\prime}
+Y_E \overline{L}_L \tilde{H}_2 E_R^\prime
+Y_E^\prime \overline{L}_R \tilde{H}_2 E_L^{\prime}\nn \\
&&
 +\Big[M \overline{L}_L L_R + M_E \overline{E}_L^{\prime} E_R +  M_D \overline{D}_L^{\prime} D_R + {\rm h.c.}\Big].
\end{eqnarray}

The mass matrix in the basis of $(E, E')$ is 
\begin{equation}
{\cal M}_{E} = \left( \begin{array}{cc}  M & \tfrac{1}{\sqrt{2}} \,Y_{E} v_{2}  \\
 \tfrac{1}{\sqrt{2}} \,Y_E^\prime v_2 & M_E \end{array} \right)\,.
\end{equation}
We have similar form of ${\cal M}_D$ by changing $Y_E^{(\prime)} \to Y_D^{(\prime)}, v_2 \to v_1, M_E \to M_D$. 
We focus on the no-mixing case,
which is possible if $M_E \gg M, Y_E v_2$ and $M_D \gg M, Y_D v_1$.
Then the light masses of $E$ and $D$ are degenerate 
as $M_{E_1}=M_{D_1}=M$.
The heavy masses are $M_{E_2}=M_E$ and $M_{D_2}=M_D$,
which suppresses the contribution from $E_2$ and $D_2$.
We do not consider the mass $M$ below $M_\Phi/2$ 
since the new decay channels of $H/A\to E\bar E/ D\bar D$
raise 
the total width quickly.
We also assume that $Y_E=Y_E^\prime$ and $Y_D=Y_D^\prime$ for simplicity. 

The Yukawa terms for the VLLs in the mass eigenstate basis become
\begin{eqnarray}
-{\cal L}_{\rm Yukawa} &=&
-\frac{1}{\tb} y_E H(\overline{E}_1 E_1 + \overline{E}_2 E_2 )
+\tb y_D H(\overline{D}_1 D_1 + \overline{D}_2 D_2 ) \nn \\
&&~-i\frac{1}{\tb} y_E A(\overline{E}_1 \gamma_5 E_1 + \overline{E}_2 \gamma_5 E_2 )
-i\,\tb y_D A(\overline{D}_1 \gamma_5 D_1 + \overline{D}_2 \gamma_5 D_2 ) \nn \\
&&~
+ y_E h(\overline{E}_1 E_1 + \overline{E}_2 E_2 )
+ y_D h(\overline{D}_1 D_1 + \overline{D}_2 D_2 ) \,,
%y_{Hu_1 u_1} = y_{Hu_2 u_2} = \frac{1}{\tb}\frac{(M^U_2-M^U_1)}{2v}
\end{eqnarray}
where
$
y_E ={\sb} Y_E/{\sqrt{2}}$ and $
y_D = {\cb}Y_D/{\sqrt{2}} $.

The partial decay widths of $\Phi=h,H,A$  
in the VLL-2HDM
are
\begin{eqnarray}
\label{gamgg}
\Gamma(\Phi\to gg) &=& \frac{G_F \alpha_s^2 M_{\Phi}^3}{64\sqrt{2}{\pi^3}}
\left| \sum_q {\hat y}^\Phi_q A^\Phi_{1/2}\big(\tau_q\big)
%+{\cal A}^{\Phi}_{gg,\rm VLF}
\right|^2 \,,\\ \nn
\Gamma(\Phi\to \rr) &=& \frac{G_F \alpha_e^2 M_{\Phi}^3}{128\sqrt{2}{\pi^3}}
\left| \sum_q {\hat y}^\Phi_q N_c Q_q^2 A^\Phi_{1/2}\big(\tau_q\big)
+\sum_\ell {\hat y}^\Phi_\ell Q_\ell^2 A^\Phi_{1/2}\big(\tau_\ell\big)
%+A_1^\Phi(\tau_W)
+{\cal A}^{\Phi}_{\rr,\rm VLL}
\right|^2,
\end{eqnarray}
where $\tau_f = M_\Phi^2/(4m_f^2)$,
the relative Yukawa couplings normalized by the SM values are 
${\hat y}^{h}_{t,b,\tau}=1$,
${\hat y}^{H,A}_t = \mp 1/\tb$ and 
${\hat y}^{H,A}_{b,\tau} = \tb$ for Type II
in the aligned 2HDM,
and the loop functions $A^{H/A}_{1,1/2}(\tau)$  are 
referred to Ref.~\cite{Djouadi:2005gi}.
The VLL contributions ${\cal A}^{\Phi}_{\rr,\rm VLL}$ 
in Eq.~(\ref{gamgg}) are given as
\begin{eqnarray}
%{\cal A}^{\Phi}_{gg,\rm VLL} &=& \sum_{\rm VL}\sum_{i=1,2}
%\bigg[ \frac{ {\hat y}^\Phi_t y_U v}{M_{U_i}}A^\Phi_{1/2}(\tau_{U_i})
%+ \frac{{\hat y}^\Phi_b y_D v}{M_{D_i}} A^\Phi_{1/2}(\tau_{D_i}) \bigg]\,,\nn \\
{\cal A}^{\Phi}_{\rr,\rm VLL} &=& \sum_{\rm VLL}\sum_{i=1,2}
\bigg[ Q_{E_i}^2 \frac{ {\hat y}^\Phi_t y_E v}{M_{E_i}} A^\Phi_{1/2}(\tau_{E_i}) + Q_{D_i}^2 \frac{ {\hat y}^\Phi_b y_D v}{M_{D_i}} A^\Phi_{1/2}(\tau_{D_i}) \bigg]\,.
\end{eqnarray}
In order to greatly enhance the $H/A\to \rr$ partial decay width through VLL loop one needs multiple number of VLL families. 
In the following analysis we introduce 3 VLL families.
We vary 
 $M$ from $375\GeV$ to $600\GeV$
 and $y_{E,D}$ from $-4\pi$ to $4\pi$. 

The final comment is on the constraint from the Higgs precision data. As shown Eq.~(\ref{gamgg}), the VLL loop also contributes to $h\to \gamma\gamma$,
which is already very limited by the 8 TeV LHC data. 
If two  Yukawa couplings $y_D$ and $y_E$
are tuned as
\begin{eqnarray}
y_D = - \frac{Q_E^2}{Q_D^2} y_E = -0.25 y_E\,,
\end{eqnarray}
new contribution to the Higgs precision data vanishes if
 $E$ and $D$ are  degenerate in mass $M$. 
 If $\tb=1$, the cancellation 
 of the VLL contributions to $h\to \rr$ equally happens to the $A\to \rr$ decay.
 Since the $A$ diphoton signal is 
 usually larger than the $H$ one if no cancellation occurs,
 we choose $\tb=0.7$ in the analysis.
 Other exclusion limits from $Z\gamma$~\cite{Zr}, $b\bar{b}$~\cite{bb}, 
 $\tau^+\tau^-$~\cite{tautau}, and $jj$~\cite{jj} channels
 at the 8 TeV LHC are satisfied in the parameter space under consideration.

%%%%%
\subsection{Results -- VLL-2HDM Model} \label{sec:result-2hdm}
We first discuss the total widths of $H$ and $A$,
both of which are dominated by the $t\bar{t}$ decay channel.
Using the running top quark mass $m_t(\mu=750\GeV) = 147\GeV$~\cite{Chetyrkin:2000yt},
we have $\Gamma_{H(A)} = 46(58)\GeV$. Since the degenerate $H$ and $A$ do not interfere, we treat them as BW peaks.
We perform a minimum $\chi^2$ analysis (see \Sec{subsec:best-fit}) and find the best-fit signal rates to the LHC 13+8 datasets
 \begin{eqnarray}
 \sigma (pp \to \Phi \to \rr ) = \left\{
 \begin{array}{c}
 6.5 \pm 2.5\fb~ (68\%{\rm CL}) \\
 6.5^{+4.5}_{-3.5}\fb ~(95\%{\rm CL})
 \end{array}
 \right.\,,
 \end{eqnarray}
which  are in agreement with Ref.~\cite{Falkowski:2015swt}.

In our scenario of VLL-2HDM 
the relative interference phase corresponds to almost imaginary interference:
\beq
\phi \simeq 
\left\{
\begin{array}{rl}
90^\circ & \hbox{ for } y_E>0; \\
-90^\circ & \hbox{ for } y_E<0. 
\end{array}
\label{eq:phi}
\right.
\eeq 
The reasons are as follows. The complex phase from the continuum background amplitude is minor~\cite{Jung:2015sna}. But the production part $gg \to H/A$ is dominated by top quark loop and the loop function generate large complex phase: $77^\circ (91^\circ)$ for $M_H(M_A)=750\GeV$. 
In addition, the decay part $H/A \to \rr$, dominated by VLL loop contribution,
is also real since $M_L >M_{H,A}/2$ in our scenario. Depending on the sign of Yukawa coupling $y_E$, the whole complex phase is changed by $\pi$. 
It maximally enhances the signal rate for $\phi\approx 90^\circ$ (constructive interference) and maximally suppress the signal rate for $\phi\approx -90^\circ$ (destructive interference). 

\begin{figure}[t] \centering
\includegraphics[width=0.49\textwidth]{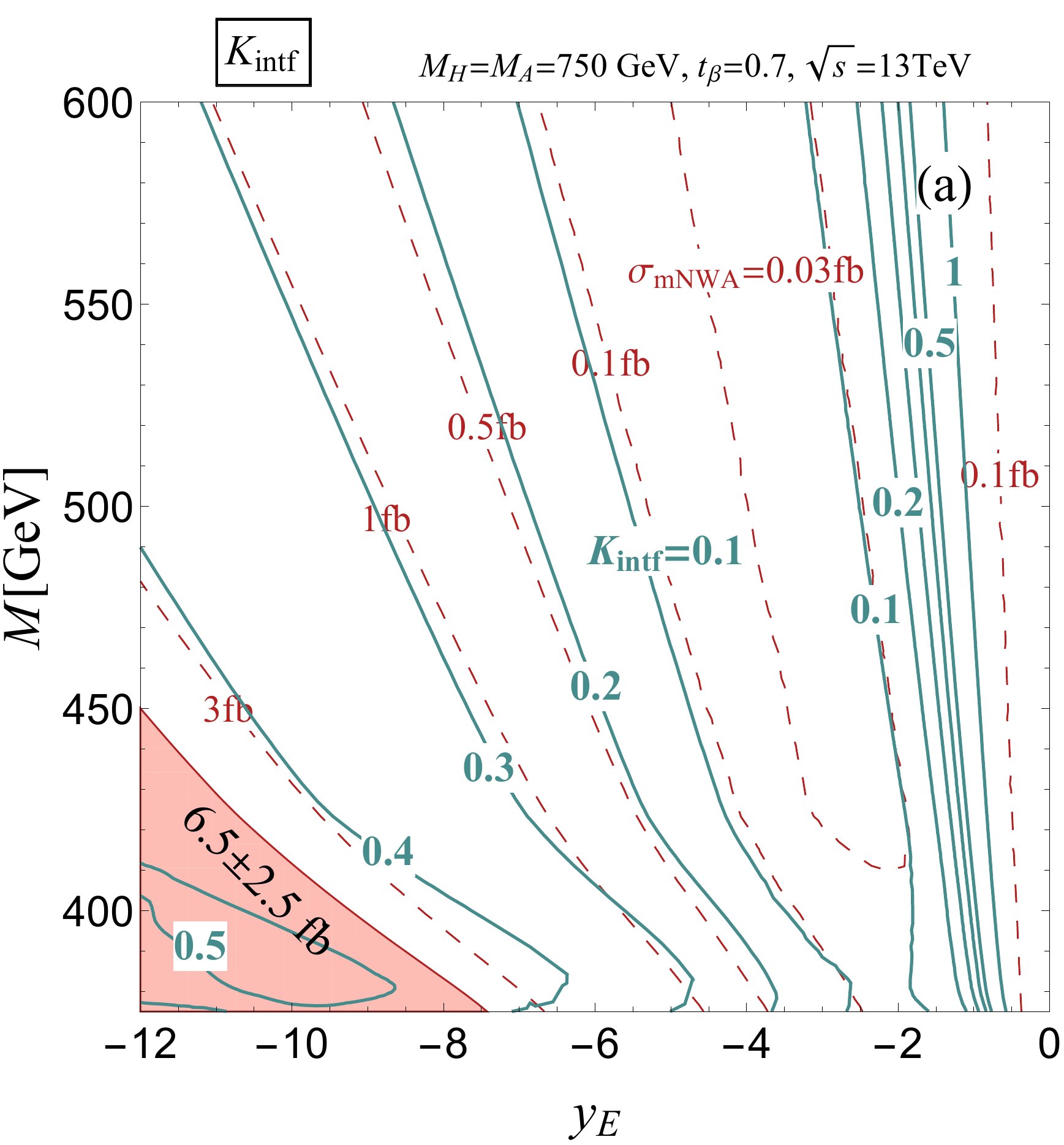}
\includegraphics[width=0.49\textwidth]{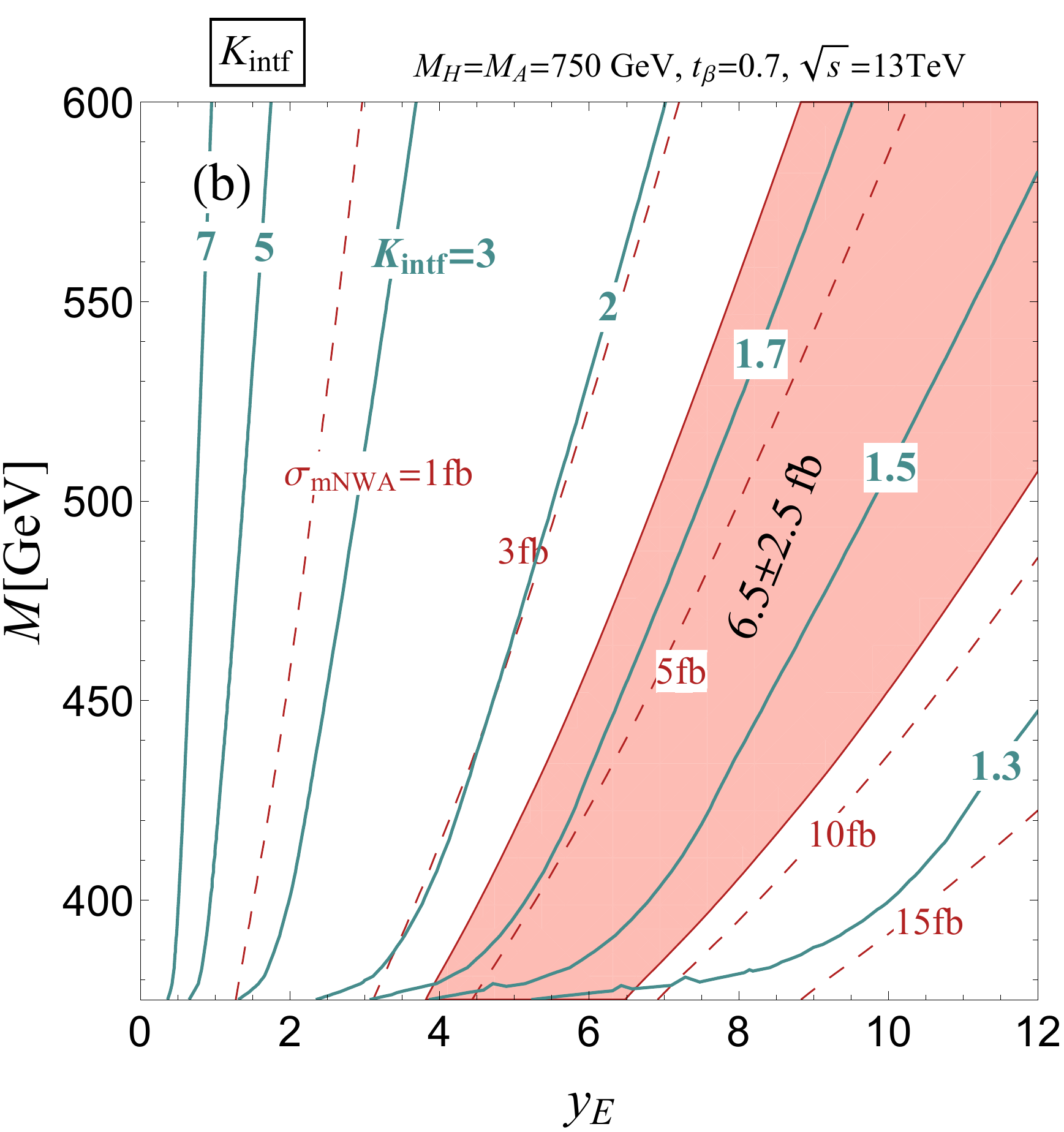}\\
\includegraphics[width=0.49\textwidth]{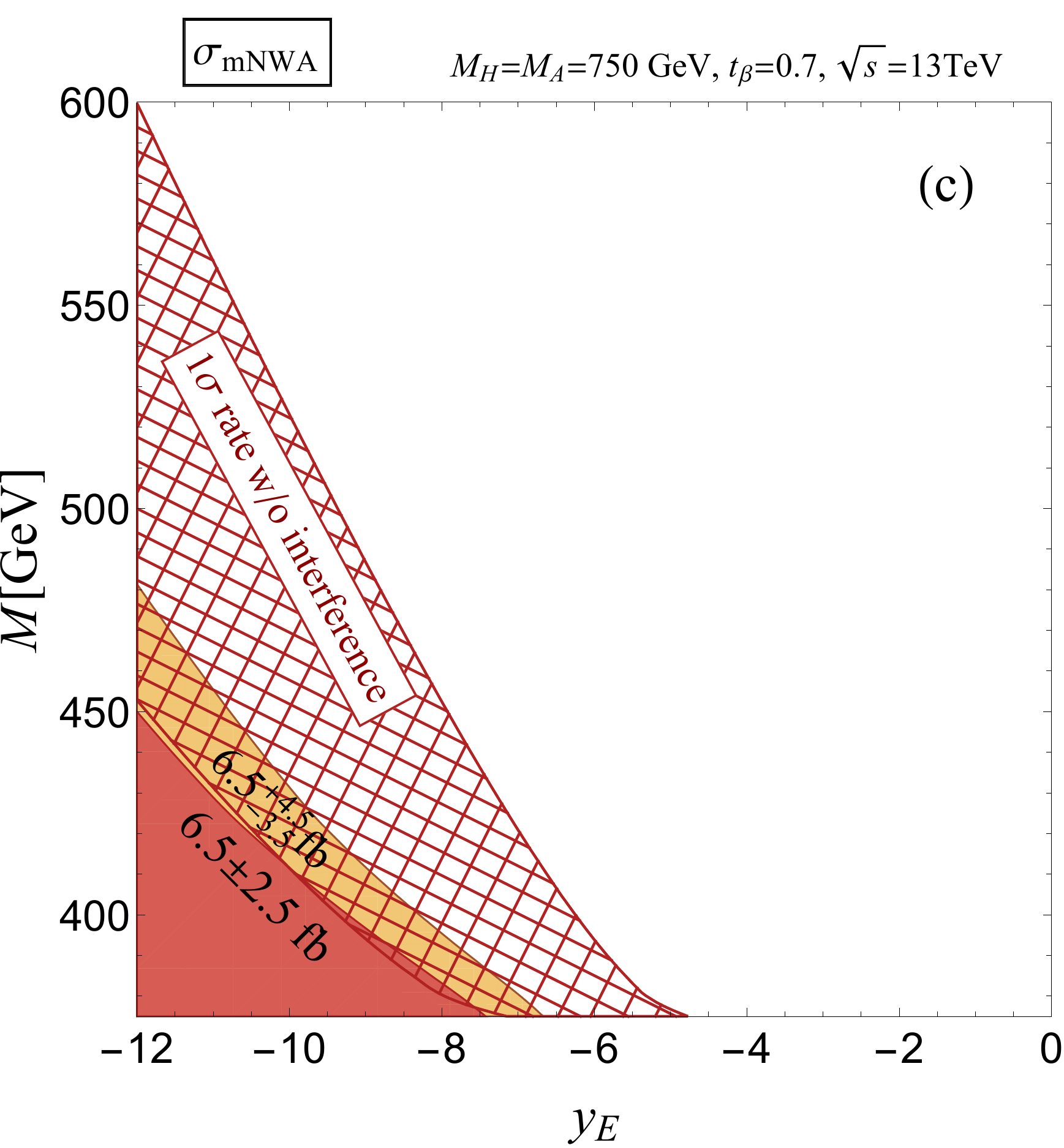}
\includegraphics[width=0.49\textwidth]{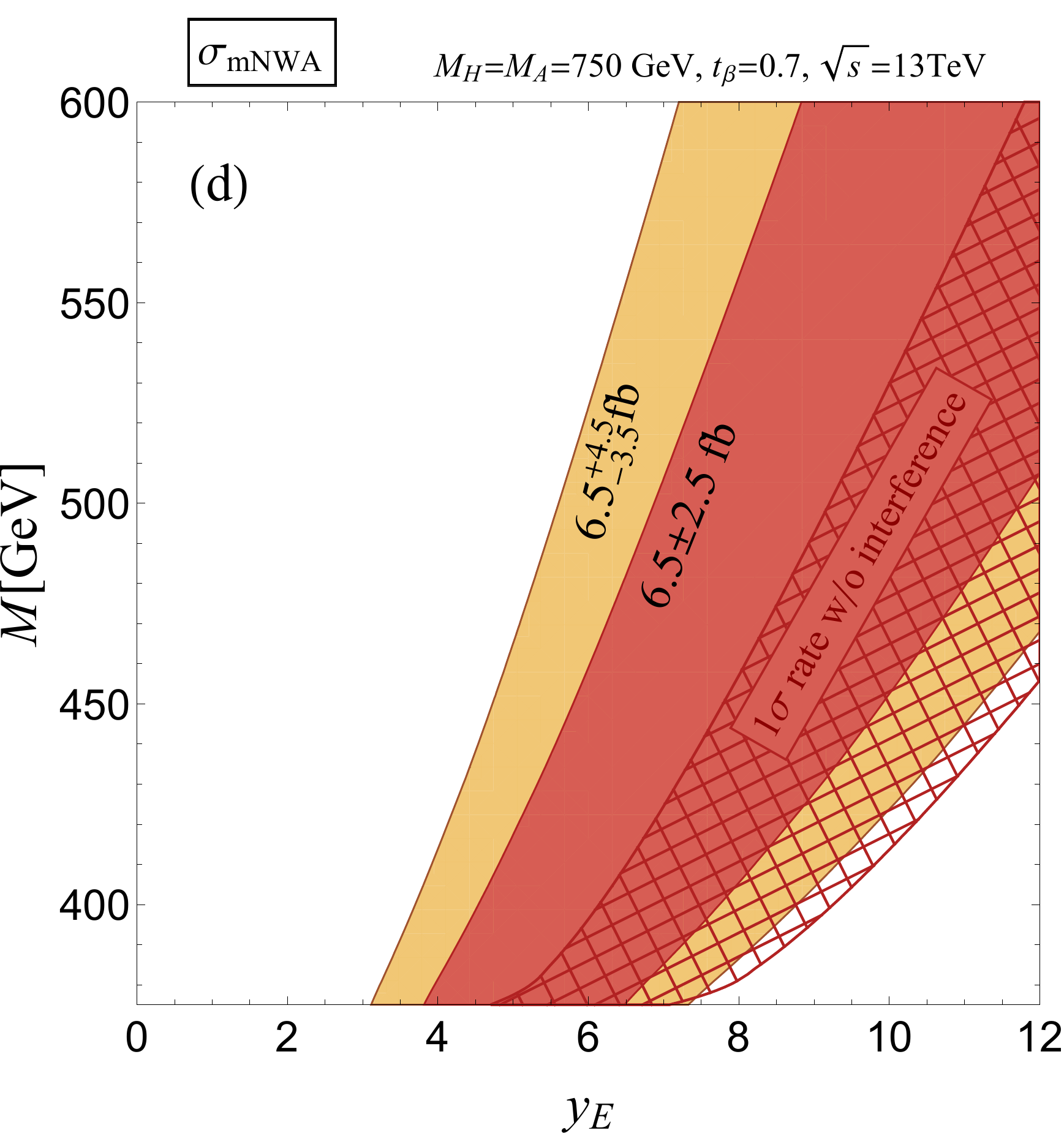}\\
\caption{\label{fig:2hdm:final}
({\bf Upper}): Contour plots for $K_{\rm intf}$ (solid lines) and $\sigma_{\rm mNWA}$ (dashed line) in the $(y_E, M)$ plane of VLL-2HDM model. ({\bf Lower}): The 68\% CL($1\sigma$, darker oranger) and 95\% CL (lighter orange) best-fit regions are shown. For comparison, the 68\% CL results without interferences accounted for are also shown as hatched regions. ({\bf Left}): $y_E<0$ induces signal suppression. ({\bf Right}): $y_E>0$ induces signal enhancement.
 \bigskip \bigskip
}
\end{figure}

Figure \ref{fig:2hdm:final}
shows our results in the parameter space $(y_E,M)$ for the VLL-2HDM:
the $K_{\rm intf}$ in \Eq{eq:Kintf} (upper panels)
and the allowed parameter space by the 750 GeV diphoton excess data 
(lower panels).
It is of great interest that quite large interference effects 
(large $K_{\rm intf}-1$) appear around the measured total signal rate,
as shown in
Figs.~\ref{fig:2hdm:final}(a) and (b).
For $y_E>0$, $K_{\rm intf}>1$ and thus constructive interference occurs:
the interference effect can make even factor 2 for the $3\fb$ total rate. 
Within the 68\% CL best-fit signal rate the interference effect ranges from $40\%$ to $80\%$ when $y_E>0$.
For $y_E<0$, $K_{\rm intf}<1$ so that destructive interference occurs:
in order to explain the signal rate, 
we need quite large magnitude of $y_E$
and thus very limited parameter space is allowed.
Figures~\ref{fig:2hdm:final}(c) and (d) show that the allowed parameter space significantly change by including the interference effect. For comparison,
we show the allowed parameter region without including interference effects.
With positive $y_E$ and $M=400\GeV$, for example,
required $y_E$ for the signal rate $6.5\fb$ 
is reduced from $\sim 7.5$ to $\sim 5.5$
by including the interference effects. 
Equivalently, the change of required number of VLL family is from 3 to 4.
In all, interference effects have significant implications
on the underlying physics.

\begin{table}[h]
\begin{center}
\begin{tabular}{|cc|cccc|cc|}
\hline
\hline
~ $M[\GeV]$ ~& $y_E$ ~&~
$\phi^H[^\circ]$ ~&~ $\phi^A[^\circ]$ ~&~ $K_{\rm intf}^H$ ~&~ $K_{\rm intf}^A$
~&~ $K_{\rm intf}^{H+A}$ ~&~ $\sigma_{\rm mNWA}^{H+A}[\fb]$ \\
\hline
457 ~& 2 ~&~ 99 ~&~ 123 ~&~ 2.6 ~&~ 5.9  ~&~ 3.5 ~&~ 1 \\
413 ~& 4 ~&~ 93 ~&~ 108 ~&~ 1.6 ~&~ 3.0  ~&~ 2.0 ~&~ 3 \\
400 ~& 6 ~&~ 91 ~&~ 104 ~&~ 1.3 ~&~ 2.1   ~&~ 1.6 ~&~ 6 \\
385 ~&$-5$ ~&~ $-96$~&~ $-88$ ~&~ 0.38 ~&~ 0.20 ~&~ 0.32 ~&~ 1 \\
395 ~&$-8$ ~&~ $-95$~&~ $-86$~&~ 0.54 ~&~ 0.21  ~&~ 0.43 ~&~ 3 \\
\hline
\hline
\end{tabular}
\caption{\label{tab:benchmark:result}
Numerical values for $\phi$, $K_{\rm intf}$, and $\sigma_{\rm mNWA}$ for $H$, $A$ and the total. The benchmark parameter points are chosen to yield total signal rates of $1,3,6\fb$. VLL-2HDM model.
}
\end{center}
\end{table}

In Table~\ref{tab:benchmark:result},
we present the numerical values for $\phi$, $K_{\rm intf}$, and $\sigma_{\rm mNWA}$. The benchmark parameter points are chosen to yield total signal rates of $1,3,6\fb$.
In order to see the individual interference effects,
we show $\phi$ and $K_{\rm intf}$ for $H$ and $A$ separately.
For both $H$ and $A$,
the relative interference phase is about $\pm 90^\circ$:
almost purely imaginary interference occurs.
$K_{\rm intf}^H$ and $K_{\rm intf}^A$
show that the interference effects are larger for $A$ than for $H$.
This is attribute to different loop functions
and thus different $M_\Phi$ positions for the vanishing real part of the 
corresponding loop functions.
One crucial result is that
the interference effects become larger with decreasing signal rate 
$\sigma^{H+A}_{\rm mNWA}$.
For 1 fb signal rate, for example, 
the enhancement factor due to the interference can be 
as large as a factor of three.
It is very possible that the current signal rate is fluctuated up and
the future precision measurement may lead to lower signal rate.
Then the interference effects become crucial.

%%%%%%%%%%%%%%%%%%%%
\section{Conclusions and Discussions} 
\label{sec:conclusions}

We have investigated the impacts of the resonance-continuum interference in  the $gg \to \gamma\gamma$ process on the the recently observed 750 GeV diphoton excess. 
The two most important interference effects -- maximal signal enhancement from the purely imaginary-part interference and maximal shape distortion from the purely real-part interference -- have been studied in two benchmark models. First, a CP-odd singlet scalar (extended with vector-like fermions) represents the purely real-part interference case, and it predicts that the 68\%(95\%) CL best-fit mass range shifts by 1--4 (any ${\cal O}(1)$) GeV. The shift is expected to be larger with a weaker coupling parameter space, which will be more preferred if the excess rate decreases in the future. Second, the heavy Higgs bosons in the two-Higgs-doublet-model (extended with vector-like leptons) represent the purely imaginary-part interference case, and the diphoton resonance signal is found to be enhanced or suppressed by a factor of 2(1.6) for 3(6) fb signal rate. Again, the effect is bigger for a weaker coupling parameter space.

Although our results are obtained with benchmark models, any scalar resonance in the $gg\to \gamma \gamma$ process
with similar widths and total rates
would exhibit similar sizes of interference effects; and the relative phase $\phi$ between the resonance and the continuum will determine the type of interference effects. 
For the given diphoton rate and the phase $\phi$, the total 
width is the most important parameter. If the width is much smaller than the current resolution $\sim {\cal O}(1)$ GeV, the real-part interference will cancel out and the imaginary-part interference will be small in proportion to the width. If a resonance is very broad, a careful study of resonance shape including its $m_{\gamma \gamma}$-dependence shall be carried out, regardless of interference effects, based on our formalism and method presented in this paper.

The future precision shape measurements and interpretations taking into account the resonance-continuum interference can provide important information and consistency check of a new resonance. One can not only test a BW resonance hypothesis but also measure $\phi$ (and the rate, mass, width). Such precision observables, in particular $\phi$, can subsequently be interpreted in terms of the properties of new particles running in loops. Remarkably, if any noticeable deviations from a BW shape can be fit well with the real-part interference, this would just be another convincing evidence of a new resonance.

%%%%%%%
\acknowledgments
The work of SJ is supported by the US Department of Energy under contract DE-AC02-76SF00515.
The work of JS is supported by NRF-2013R1A1A2061331.
The work of YWY is supported by NRF-2012R1A2A1A01006053.
We thank KIAS Center for Advanced Computation for providing computing resources.

%%%%%%%%%%%%%%%%%%%%%%

\end{document}